\documentclass[12pt]{iopart}
\expandafter\let\csname equation*\endcsname\relax
\expandafter\let\csname endequation*\endcsname\relax 

\usepackage{amsmath}
\usepackage{units}

\usepackage{braket}

\usepackage{graphicx}

\usepackage{media9}

\usepackage{cite}

\begin{document}
\title[Ultrafast spin dynamics in inhomogeneous systems]{Ultrafast spin dynamics in inhomogeneous systems:\\ a density-matrix approach applied to Co/Cu interfaces}

\author{F T\"opler, J Henk, I Mertig}
 
\address{Martin Luther University Halle-Wittenberg, Institute of Physics, 06099 Halle, Germany}

\ead{franziska.toepler@physik.uni-halle.de}

\date{\today}

\begin{abstract}
Ultrafast spin dynamics on femto- to picosecond timescales is simulated within a density-operator approach for a Co/Cu bilayer. The electronic structure is represented in a tight-binding form; during the evolution of the density operator, optical excitation by a femtosecond laser pulse, coupling to a bosonic bath as well as dephasing are taken into account. Our simulations corroborate the importance of interfaces for ultrafast transport phenomena and demagnetisation processes. Moreover, we establish a reflow from Cu $d$ orbitals across the interface into Co $d$ orbitals, which shows up prominently in the mean occupation numbers. On top of this, this refilling manifests itself as a minority-spin current proceeding several layers into the Cu region. The present study suggests that the approach captures essential ultrafast phenomena and provides insight into microscopic processes.
\end{abstract}


\submitto{\NJP}
\maketitle

\section{Introduction} \label{sec:Intro}
The field of spin-electronics comprises a variety of effects that often show up in inhomogeneous systems and as responses to a static cause. In recent years, however, the attention has shifted to ultrafast processes which, for example, are pushed by femtosecond laser pulses or by electromagnetic terahertz radiation. Understanding the fundamentals of ultrafast spin dynamics requires theories which allow to identify the essential mechanisms for, e.\,g., switching magnetisation or generating spin-polarised currents, provide intuitive but detailed pictures of these processes, and are able to reproduce experimental results.

Ultrafast spin dynamics in inhomogeneous systems covers time scales from femto- to picoseconds. On the femtosecond timescale time-dependent density-functional theory (TDDFT) offers the perhaps most fundamental and detailed description \cite{krieger_laser-induced_2015}. Since such computations are very demanding, they are presently limited to a short duration (a few ten femtoseconds) and a small sample size (a few unit cells). On top of this, the electron system is not coupled to thermodynamic baths (e.\,g., phonons or magnons). 

T\"ows and Pastor developed a many-particle theory of ultrafast demagnetisation in ferromagnetic transition metals \cite{Toews2014,Toews2015}, in which spin-orbit coupling and a laser pulse are considered. On top of this, the intra-atomic Coulomb interaction is taken into account, making analytical or numerical solutions very demanding. An application to Ni showed that demagnetisation effects are short-ranged \cite{Toews2015}, allowing to use small clusters in configuration space.

On the picosecond timescale, at which thermalisation takes place, the subsystems of electrons, phonons, and magnons may be described by their individual temperatures that are obtained from rate equations, as in the three-temperature model \cite{koopmans_explaining_2010}. The detailed geometry of a sample, for example inhomogeneities, is usually regarded as side issue.

The above calls for a versatile theoretical single-particle approach that covers all timescales on equal footing but does not suffer from restrictions on a sample's geometry and size. On top of this, it should account for both the excitation of the electron system by a laser pulse and the coupling of the electron system to  thermodynamic baths in order to achieve thermalisation. In this paper we report on a theoretical method which aims at fulfilling the previously mentioned objectives. The starting basis is a tight-binding description of the electronic structure in configuration space which is not only versatile but as well provides an intuitive interpretation of the results. As in TDDFT simulations, we obtain mean occupation numbers and currents by evolving the one-particle density operator in time. The occupation numbers and currents are fully resolved with respect to time, site, orbital, and spin, thereby allowing a detailed analysis. Thermalisation is achieved by coupling the electron system to a bosonic bath. As an application we choose spin dynamics at Co/Cu interfaces which due to their epitaxial growth serve as paradigm system in spin-electronics. 

Some theories of transport in a ferromagnetic/nonmagnetic sample concentrate on majority-spin electrons, for example superdiffusive spin transport \cite{battiato_superdiffusive_2010,battiato_superdiffusive_2013}. Strong asymmetries of spin-dependent relaxation times and transition probabilities explain  that after the laser excitation mostly hot majority-spin electrons traverse the interface and propagate rapidly from the ferromagnetic into the nonmagnetic region of the sample. This scenario is corroborated by experimental results on Co/Cu systems that show that demagnetisation is initially governed by spin-polarized transport starting at the interface \cite{wieczorek_separation_2015} (for Fe/Au see Refs.~\cite{Melnikov2011} and~\cite{nenno_particle--cell_2018}).
 
The above sequence of events is complemented by a backflow mechanism \cite{chen_competing_2019,borchert_manipulation_2020}: empty minority-spin Co $d$ orbitals become occupied by Cu $d$ electrons from sites adjacent to the Co/Cu interface. The Co $d$ orbitals can be located above or below the chemical potential. In the latter case they have to be depleted by a laser excitation before being refilled. Chen \etal explained this `reflow' by resonant optical excitation by the $\unit[1.5]{eV}$ laser pump pulse; it is in line with the high density of states of both Co and Cu in the relevant energy range. The reflow mechanism is reminiscent to optically induced inter-site spin transfer (OISTR; Ref.~\cite{dewhurst_laser-induced_2018}).

The reflow scenario in Co/Cu is fully confirmed by our simulations. As will be demonstrated in this paper, we identify which orbitals are essential for the above mechanism. Furthermore, the currents triggered by the reflow are investigated for times considerably after the pulse, thereby going beyond the TDDFT simulations by Chen \etal \cite{chen_competing_2019}. Space-time maps of the associated spin-resolved currents reveal how deep the reflow penetrates into the Cu region of the samples.

This paper, whose main purpose is to introduce to the approach and to demonstrate its key capabilities, is organised as follows. In \Sref{sec:TheoreticalAspects} we sketch our theoretical approach by addressing sample geometry (\Sref{sec:Geometry}), electronic structure (\Sref{sec:ElectronicStructure}), and evolution of the density operator (\Sref{sec:EvolutionDM}). In \Sref{sec:Applications} we report on selected features of spin dynamics at Co/Cu interfaces (\Sref{sec:SetupCoCu}): occupation numbers and demagnetisation (\Sref{sec:OccNumDemag}) as well as spin transfer across the interface (\Sref{sec:SpinTransfer}). An outlook is given in \Sref{sec:Outlook}, while further details are comprised in an appendix.

\section{Theoretical aspects} \label{sec:TheoreticalAspects}
\subsection{Sample geometry} \label{sec:Geometry}
A sample is represented as a cluster of atoms in configuration space. In order to be flexible, a cluster is built from blocks, each of which contains identical unit cells; the unit cells of the individual blocks may differ from each other. This approach allows to construct samples in one, two or three spatial dimensions, for example chains, heterostructures or homogeneous systems with defects (e.\,g., vacancies). A typical cluster's shape is a parallelepiped.

\subsection{Electronic structure} \label{sec:ElectronicStructure}
The electronic structure of a beforehand defined cluster is described within the Slater-Koster formulation of the tight-binding (TB) method \cite{Slater1954}. For each site (atom) we consider exchange-split $s$, $p$, and $d$ orbitals and take into account spin-orbit coupling. The spin quantisation axis is the $z$ axis in the global frame. The tight-binding parameters are either taken from literature or computed from first principles. Furthermore, the boundary conditions (open or closed) are specified for each edge direction of the cluster parallelepiped.

The result of the above setup is the time-independent Hamiltonian matrix $\mathsf{H}_{0}$ whose eigenvectors enter the evolution of the one-particle density matrix $\mathsf{P}(t)$; see \Sref{sec:EvolutionDM} below. Recall that the TB parameters do not depend on time, which is a reasonable assumption as long as any time-dependent perturbation is weak. This implies that a so-called transient band structure \cite{Eich2017} can show up in a simulation only as variation of the mean occupation numbers of the eigenstates of $\hat{H}_{0}$; the eigenstates and -energies themselves are not affected.

The spin dynamics is analyzed in terms of the $\hat{H}_{0}$ eigenstates $\ket{n}$ (termed `eigenstate basis' in this paper) or in the `site-orbital basis' of site-dependent orbitals $\ket{k, \alpha}$ ($k$ site index, $\alpha$ multi-index comprising orbital and spin quantum numbers), as the occasion may require.

\subsection{Evolution of the density operator} \label{sec:EvolutionDM}
The evolution of the one-particle density operator $\hat{\rho}(t)$ of the electron system is determined by a laser pulse and by coupling to a bosonic bath.

A laser pulse is specified by its electric-field vector (direction of incidence with respect to the global frame, amplitude, and polarisation), the photon energy, and the shape of a Gaussian envelope function, the latter typically of a few $\unit{fs}$ width. The pulse induces optical dipole transitions between the orbitals, whose matrix elements are condensed into the perturbation matrix $\mathsf{V}(t)$; cf.\ Ref.~\cite{Toews2015}. The spin-conserving transitions obey the angular-momentum selection rules $\delta l = \pm 1$ and $\delta m = 0, \pm 1$, depending on the laser's polarisation.

The electron system is thermalised by coupling it to a fictitious bosonic bath, the latter mimicking phonons or magnons. The coupling is described by Lindblad superoperators \cite{lindblad_generators_1976, Pershin2008} that determine how rapidly the laser-excited electron system will relax toward thermal equilibrium. The Lindblad superoperator 
\begin{align}
    \mathcal{L}[\hat{F}](\hat{\rho}) & \equiv \hat{F} \hat{\rho} \hat{F}^{\dagger} - \frac{1}{2} \left( \hat{F}^{\dagger} \hat{F} \hat{\rho} +  \hat{\rho} \hat{F}^{\dagger} \hat{F} \right)
    \label{eq:SuperOp}
\end{align}
of an operator $\hat{F}$ acts on the density operator $\hat{\rho}$ and accounts for energy transfer from the electron system to the bath and vice versa; it describes also dephasing. The number of electrons is conserved since $\mathcal{L}[\hat{F}](\hat{\rho})$ is traceless.

A jump operator $\hat{F}_{nm} \equiv \ket{n} \bra{m}$ mediates an inelastic transition from the $\hat{H}_{0}$ eigenstate $\ket{m}$ with energy $\varepsilon_{m}$ into the eigenstate $\ket{n}$ with energy $\varepsilon_{n}$, thereby transferring an energy of $\Delta \varepsilon$ from the electron system into the bath ($\varepsilon_{m} > \varepsilon_{n}$) or vice versa ($\varepsilon_{n} > \varepsilon_{m}$). The Lindbladian $\mathcal{L}[\hat{F}_{nm}](\hat{\rho}(t))$ for $\hat{F} = \hat{F}_{nm}$ at time $t$,
\begin{align}
    \hat{L}_{nm}(\hat{\rho}(t))
    & \equiv \mathcal{L}[\hat{F}_{nm}](\hat{\rho}(t))
    = \ket{n} \rho_{mm}(t) \bra{n} 
    - \frac{1}{2} \sum_{i} \left( \ket{m} \rho_{mi}(t) \bra{i} + \ket{i} \rho_{im}(t) \bra{m} \right),
\end{align}
follows from \Eref{eq:SuperOp}; $\rho_{nm}(t) \equiv \braket{n | \hat{\rho}(t) | m}$. The total Lindbladian 
\begin{align}
    \hat{L}(\hat{\rho}(t)) & \equiv \sum_{nm} \gamma_{nm}(t) \, \hat{L}_{nm}(\hat{\rho}(t))
    \label{eq:totalLinbdbladian}
\end{align}
is then obtained by a weighted sum over all $\hat{L}_{nm}$.

If $n \neq m$, the strengths $\gamma_{nm}$ account for the Bose-Einstein distribution function $f_{\mathrm{BE}}$ of the bath as well as for spin-conserving (sc) and spin-flip (sf) transitions. For the latter we introduce transition rates $\gamma_{\mathrm{sc}}$ and $\gamma_{\mathrm{sf}}$, respectively.

If $n = m$, the Lindblad operators $\hat{L}_{nn}$ are energy-conserving and do not represent transitions. Instead they cause pure dephasing by reducing the off-diagonal elements of $\hat{\rho}$; their diagonal elements vanish. The dephasing rate $\gamma_{\mathrm{dp}}$ determines the timescale on which modulations of the mean occupation numbers are quenched. 

Summarizing, $\gamma_{nm}(t)$ in \Eref{eq:totalLinbdbladian} takes the form
\begin{align}
\gamma_{nm}(t) & = 
\begin{cases}
\gamma_{\cdot} \, \pi_{nm}(t) \left[ f_{\mathrm{BE}}(\Delta\varepsilon, \mu, T) + 1 \right] & \text{de-excitation}, n \not= m \\
\gamma_{\cdot}\,\pi_{nm}(t) \, f_{\mathrm{BE}}(\Delta\varepsilon, \mu, T) & \text{excitation}, n \not= m \\
    \gamma_{\mathrm{dp}} & \text{dephasing}, n = m
    \end{cases},
    \label{eq:LindParams}
\end{align}
with $\cdot$ being a placeholder for either `sc' or `sf'. The factors $\pi_{nm}(t) = \rho_{mm}(t) \, \left[ 1 - \rho_{nn}(t) \right]$ account for the Pauli exclusion principle and ensure $0 \leq \rho_{nn}(t)\le 1$ for each eigenstate and at all times \cite{Coleman1960,Chakraborty2014,Ngyuen2015,Head2015}.

Representing the operators as matrices, the evolution of the density matrix $\mathsf{P}(t)$ (with matrix elements $\rho_{nm}(t)$) is given by the Lindblad equation
\begin{align}
    \partial_{t} \mathsf{P}(t) & = \mathrm{i} \left[ \mathsf{P}(t), \mathsf{H}_{0} + \mathsf{V}(t) \right] + \mathsf{L}(\mathsf{P}(t))
    \label{eq:vonNeumann}
\end{align}
 (in Hartree atomic units). $\mathsf{P}(t)$ is not diagonal because of the perturbation $\mathsf{V}(t)$.

The outcome of the Lindblad equation are mean occupation numbers $\rho_{nn}(t)$ of the eigenstates $\ket{n}$ and -- after transformation into the site-orbital basis -- of the orbitals $\ket{k, \alpha}$, $\rho_{k \alpha, k \alpha}(t)$. The effect of the perturbation and the coupling to the bath is identified as the difference $\Delta n(t) \equiv \rho_{nn}(t) - \rho_{nn}^{(0)}$, that is the deviation of the nonequilibrium occupancy $\rho_{nn}(t)$ with respect to the one in equilibrium ($\rho_{nn}^{(0)}$, an element of $\mathsf{P}_{0}$).

Following a derivation by Mahan \cite{mahan_many-particle_2000},
the current flowing from site $l$ to site $k$ at time $t$ can be expressed as
\begin{align}
    \braket{j_{kl}(t)} & =  \frac{\mathrm{i}}{2} \operatorname{tr} \mathsf{P}_{lk}(t) \, \mathsf{H}_{kl} - \left\{ l \leftrightarrow k \right\}
    \label{eq:current},
\end{align}
in which 
\begin{align*}
\mathsf{P}_{lk}(t) & =
\begin{pmatrix}
\mathsf{p}_{lk}^{\uparrow \uparrow}(t) & \mathsf{p}_{lk}^{\uparrow \downarrow}(t) \\
\mathsf{p}_{lk}^{\downarrow \uparrow}(t) & \mathsf{p}_{lk}^{\downarrow \downarrow}(t)
\end{pmatrix},
\quad
\mathsf{H}_{kl} =
\begin{pmatrix}
\mathsf{h}_{kl}^{\uparrow \uparrow} & \mathsf{h}_{kl}^{\uparrow \downarrow} \\
\mathsf{h}_{kl}^{\downarrow \uparrow} & \mathsf{h}_{kl}^{\downarrow \downarrow}
\end{pmatrix},
\end{align*}
are blocks of the density matrix $\mathsf{P}(t)$ and the Hamilton matrix $\mathsf{H}_{0}$, respectively, both expressed in the site-orbital basis. The submatrices $\mathsf{p}_{lk}^{\sigma \sigma'}(t)$ and $\mathsf{h}_{kl}^{\sigma \sigma'}$ are indexed by orbital. For details see~\ref{sec:spincurr}.

The spin-polarised currents resolved with respect to the $\mu$-th component are given by 
\begin{align*}
\braket{j_{kl}^{\mu}} & = 
\frac{\mathrm{i}}{4} \operatorname{tr}
\mathsf{P}_{lk}
\left[ \mathsf{\Sigma}^{\mu}, \mathsf{H}_{kl} \right]_{+}
- \left\{ l \leftrightarrow k \right\}, \quad \mu = x, y, z,
\end{align*}
in which $\mathsf{\Sigma}^{\mu}$ is a block Pauli matrix and $[ \cdot,  \cdot]_{+}$ is the $+$-commutator. The $z$-resolved spin-polarised current studied in this paper reads
\begin{align*}
\braket{j_{kl}^{z}} & = 
\frac{\mathrm{i}}{2} \operatorname{tr} \left(
\mathsf{p}_{lk}^{\uparrow \uparrow} \mathsf{h}_{kl}^{\uparrow \uparrow}
-
\mathsf{p}_{lk}^{\downarrow \downarrow} \mathsf{h}_{kl}^{\downarrow \downarrow} 
\right)
-\left\{ l \leftrightarrow k\right\},
\end{align*}
that is the usual difference of spin-up and spin-down currents.

\Eref{eq:current} may be construed broadly as follows. The laser pulse produces local imbalances of occupation (that is electric polarisation), which themselves cause site-off-diagonal $\rho$ terms. The larger the imbalance and the stronger the hopping strength between the sites, the larger the current. Recall that the hopping strengths $h_{k\alpha, l \beta}$ determine the bandwidths and, thus, the group velocities of the electrons.

\section{Applications} \label{sec:Applications}
\subsection{Setup for Co/Cu systems} \label{sec:SetupCoCu}
To illustrate the approach and its capabilities we deliberately choose rather simple Co/Cu systems which have the advantage that numerical results are easily visualised and interpreted but nevertheless exhibit the relevant physical processes.

A two-dimensional cluster of 20~Co and 20~Cu atoms simulates dynamics in a thin fcc Co layer on a Cu substrate with a (001) interface. Periodicity is applied in the directions parallel to the interface ($y$ and $z$). For studying spin transfer we replaced this cluster by one with 10~Co and 30~Cu atoms.

The tight-binding parameters were taken from Ref.~\cite{papaconstantopoulos_handbook_2015} but with the on-site energies of bulk Co and Cu adjusted to achieve aligned Fermi energies. The Co-Cu hopping parameters (across the interface) are taken as arithmetic means of the respective ones of Co and Cu; we are aware that this may be regarded as too rough an approximation but recall that  similar heuristic settings worked for tunnel junctions  \cite{Mathon1999}.

The electron system is excited by a linearly polarized laser pulse with a photon energy of $\unit[1.55]{eV}$ and a width of the Gaussian envelope of $\unit[10]{fs}$.

The Lindblad parameters $\gamma_{nm}$ defined in \Eref{eq:LindParams} are chosen such as to reproduce relaxation and dephasing times in experiments: $\gamma_{\mathrm{sc}} = \unit[2\cdot 10^{-4}]{fs}^{-1}$, $\gamma_{\mathrm{sf}} = \unit[2\cdot 10^{-6}]{fs}^{-1}$, and $\gamma_{\mathrm{dp}} =\unit[5\cdot 10^{-2}]{fs}^{-1}$, which correspond to lengths of times of $\unit[5]{ps}$, $\unit[500]{ps}$ and $\unit[20]{fs}$, respectively. The results presented in this paper do not depend crucially on their actual values. The bath temperature is set to $T = \unit[300]{K}$ in all simulations.

\subsection{Mean occupation numbers and demagnetisation} \label{sec:OccNumDemag}

As addressed in the introduction (\Sref{sec:Intro}), temperature models are often used for describing spin dynamics. For example, time-resolved spectroscopic data are approximated by Fermi-Dirac distributions, yielding the evolution of both electron temperature  and chemical potential. Such an approach seems reasonable at times after a $\unit{fs}$ laser pulse, but needs to be verified for times during or short after a laser pulse. 

\begin{itemize}
    \item  Long before the laser pulse, the initial thermalisation produces a Fermi-Dirac distribution of the occupation numbers [panel~(a) in \Fref{fig:OccNumProfiles}].\footnote{\Fref{fig:OccNumProfiles} comprises selected snapshots of the evolution animated in \Fref{fig:Movie}.} Next, electron states close to the chemical potential become depopulated or populated, but  the distribution is changed slightly (not shown); this is explained by the rather weak amplitude of the laser pulse at these times.

\begin{figure}
    \centering
    \includegraphics[width = 0.96\columnwidth]{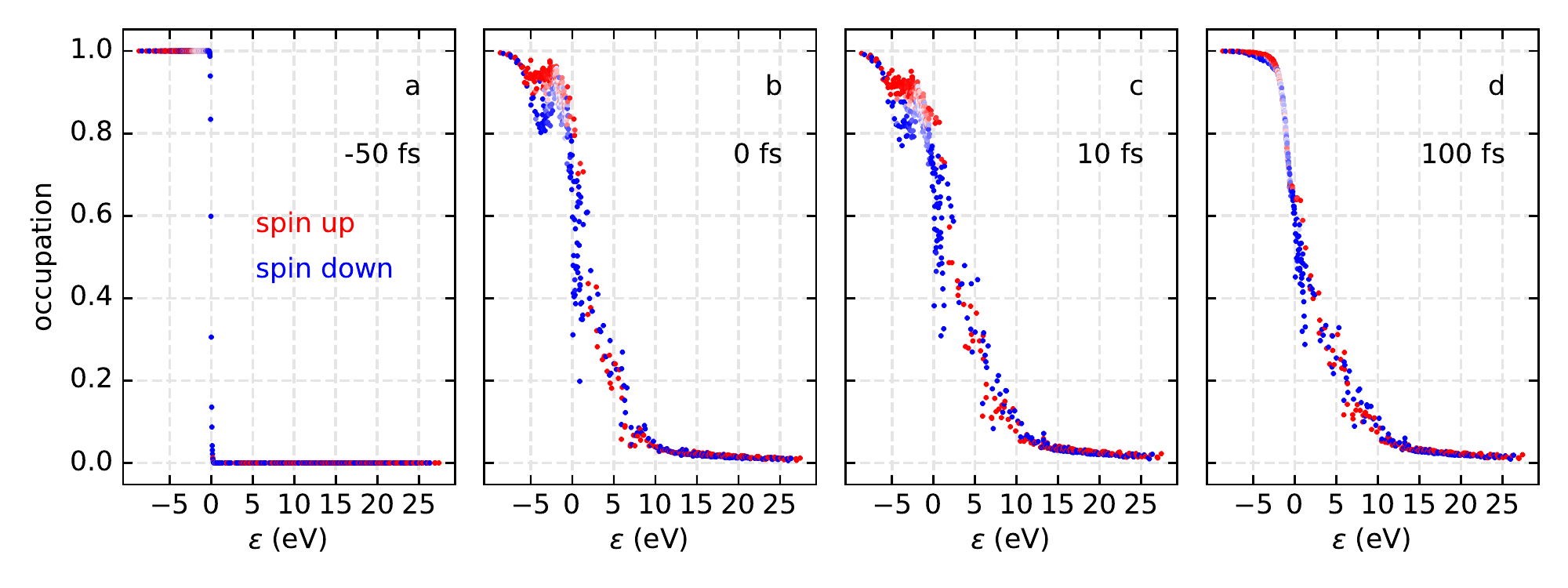}
    \caption{Energy- and spin-resolved mean occupation numbers of eigenstates of the Hamiltonian $\hat{H}_{0}$. Majority-spin (minority-spin) data are depicted as red (blue) dots for times $t = \unit[-50]{fs}$ (a), $\unit[0]{fs}$ (b), $\unit[10]{fs}$ (c), and $\unit[100]{fs}$ (d), respectively. Temperature $\unit[300]{K}$, chemical potential at $\unit[0]{eV}$, laser pulse with maximum amplitude at $\unit[0]{fs}$ and width of $\unit[10]{fs}$ (as sketched in \Fref{fig:Demag}).}
    \label{fig:OccNumProfiles}
    \end{figure}
    
    \item With increasing amplitude, also electron states well below the chemical potential~$\mu$ become strongly depopulated; simultaneously, hot electrons appear at energies far above $\mu$ [panel~(b) shows the distribution at the maximum amplitude]. The distribution is modulated in line with the frequency of the pulse (recall that dipole transitions induce both excitations and de-excitations). 
    
    \item Right after the pulse, the mean occupation numbers reproduce at first glance a Fermi-Dirac distribution for a very high temperature but with significant deviations [panel~(c)]. The population of Co $d^{\downarrow}$ orbitals appears as a drop of the broad Fermi-Dirac distribution around $\unit[+1]{eV}$. Similarly, the depopulation of Cu $d^{\downarrow}$ orbitals manifests itself as the dip at about $\unit[-3]{eV}$.
    
    \item After $\unit[100]{fs}$, that is long after the laser pulse, thermalisation makes the distribution `almost thermal' but with a sizable population of orbitals with energies above the chemical potential [compare panel~(d) with~(a)]. The depopulations below the chemical potential are considerably reduced [compare panel~(d) with~(b) and~(c)].
\end{itemize}

The laser pulse produces mainly a depopulation of minority-spin electrons below~$\mu$ and a population of hot electrons. The latter is larger, the larger the maximum amplitude of the pulse. This finding is confirmed qualitatively by occupation-number profiles computed within TDDFT for similar Co/Cu systems. 

\begin{figure}
    \centering
    \includegraphics[width = 0.96\columnwidth]{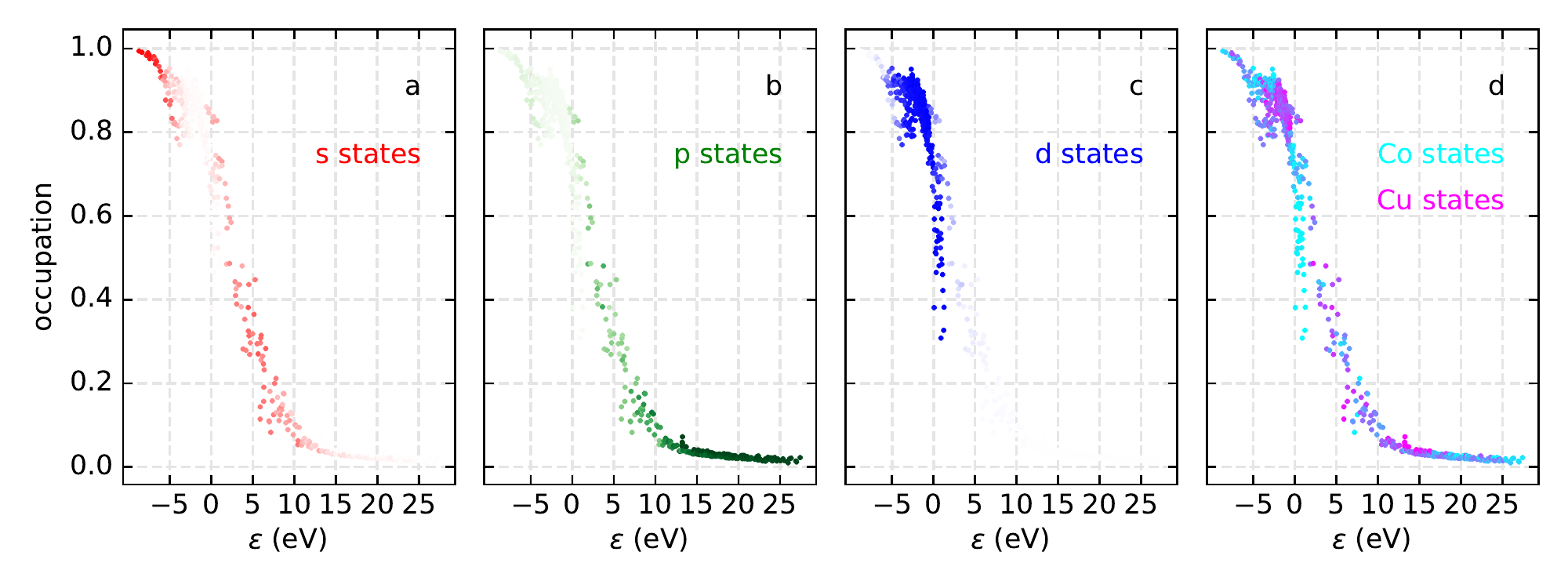}
    \caption{As \Fref{fig:OccNumProfiles} panel~c, but resolved with respect to orbital [panels a -- c: $s$ (red), $p$ (green), and $d$ (blue), respectively] and localisation [panel~d:: Cu sites (magenta) and Co sites (cyan)].}
    \label{fig:OccNumPrjct}
\end{figure}

 The pronounced structure within the distribution of minority electrons is analysed further by projecting the eigenstates onto the involved orbitals ($s$, $p$, and $d$; \Fref{fig:OccNumPrjct} panels~a -- c). Panel~c shows that this structure is composed mainly of eigenstates of $d$ orbital character. While  those states with reduced occupation below the chemical potential~$\mu$ are identified as Cu states, eigenstates with enhanced occupation exactly at and above $\mu$ exhibit Co character (panel~d). The reduction of Cu $d^\downarrow$ occupation and the enhancement of Co $d^\downarrow$ occupation already hints at a transfer of $d^\downarrow$ electrons from Cu to Co. This assumption shall be verified further in \Sref{sec:SpinTransfer}.

\begin{figure}
   \centering
    \includegraphics[width = 0.96\columnwidth]{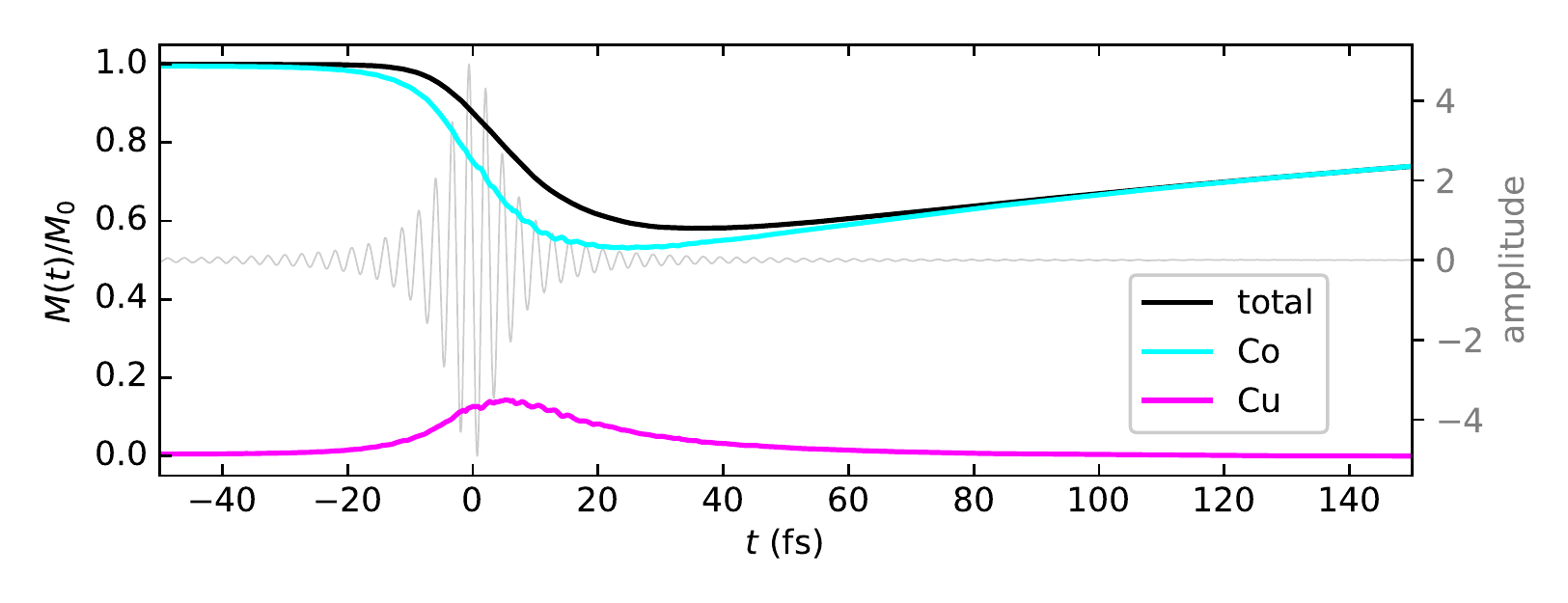}
    \caption{Evolution of the magnetisation $M(t)$ obtained from the data shown in \Fref{fig:OccNumProfiles}. The normalized magnetisation of the entire sample (black line,  `total'; $M_{0}$ magnetisation of the total sample at thermal equilibrium) is decomposed into that of the Co region (cyan, `Co') and that of the Cu region (magenta, `Cu'). The amplitude of the laser pulse (in arbitrary units; right abscissa) is represented as thin line.}
    \label{fig:Demag}
\end{figure}

The quite complicated occupation-number profiles reproduced in \Fref{fig:OccNumProfiles} translate into smooth magnetisation spectra (\Fref{fig:Demag}). The ferromagnetic part of the sample ('Co', cyan line) becomes considerable demagnetized to about $50$ per cent of its initial value already during and directly after the laser pulse. A fraction of this demagnetisation is caused by spin-conserving transfer of magnetic moment from Co into Cu (cf. \Sref{sec:SpinTransfer}). Hence the demagnetisation of the entire sample (black spectrum, `total') is weaker and follows the demagnetisation of Co  with a time delay.

The magnetisation of the Cu region (magenta in \Fref{fig:Demag}) shows a maximum at about $\unit[10]{fs}$ and likewise indicates transfer of magnetic moment from Co into Cu across the interface. A closer analysis reveals that shortly after the pulse Cu atoms gain magnetic moment which subsequently is lost due to spin-flip processes mediated by the bath or by spin-orbit coupling. We note in passing that the magnetisation dynamics in Co and Cu are very weakly modulated by the laser pulse in the range from about $t = \unit[-5]{fs}$ to about $\unit[15]{fs}$.

\subsection{Spin transfer across the interface} \label{sec:SpinTransfer}
An established picture of spin transport is that majority electrons in the magnetic region of the sample are excited by the laser pulse and then propagate across the interface into the nonmagnetic region \cite{alekhin_femtosecond_2017,battiato_superdiffusive_2010,battiato_superdiffusive_2013,nenno_particle--cell_2018}. Another mechanism is optically induced inter-site spin transfer (OISTR; Ref.~\cite{dewhurst_laser-induced_2018}). Here, a laser pulse causes a rearrangement of spin polarisation at the interface which can reverse the magnetic moments of layers. In agreement with Chen \etal \cite{Chen2019}, our simulations establish a similar mechanism: a reflow from Cu $d$ orbitals into empty Co minority-spin $d$ orbitals across the interface. To study this effect we simulated a cluster of 10~Co and 30~Cu atoms with periodicity parallel to the interface. This setup resembles a thin fcc Co layer on a Cu substrate with a (001) interface and facilitates the analysis of currents. 

\begin{figure*}
    \centering
    \includegraphics[width = 0.96\columnwidth]{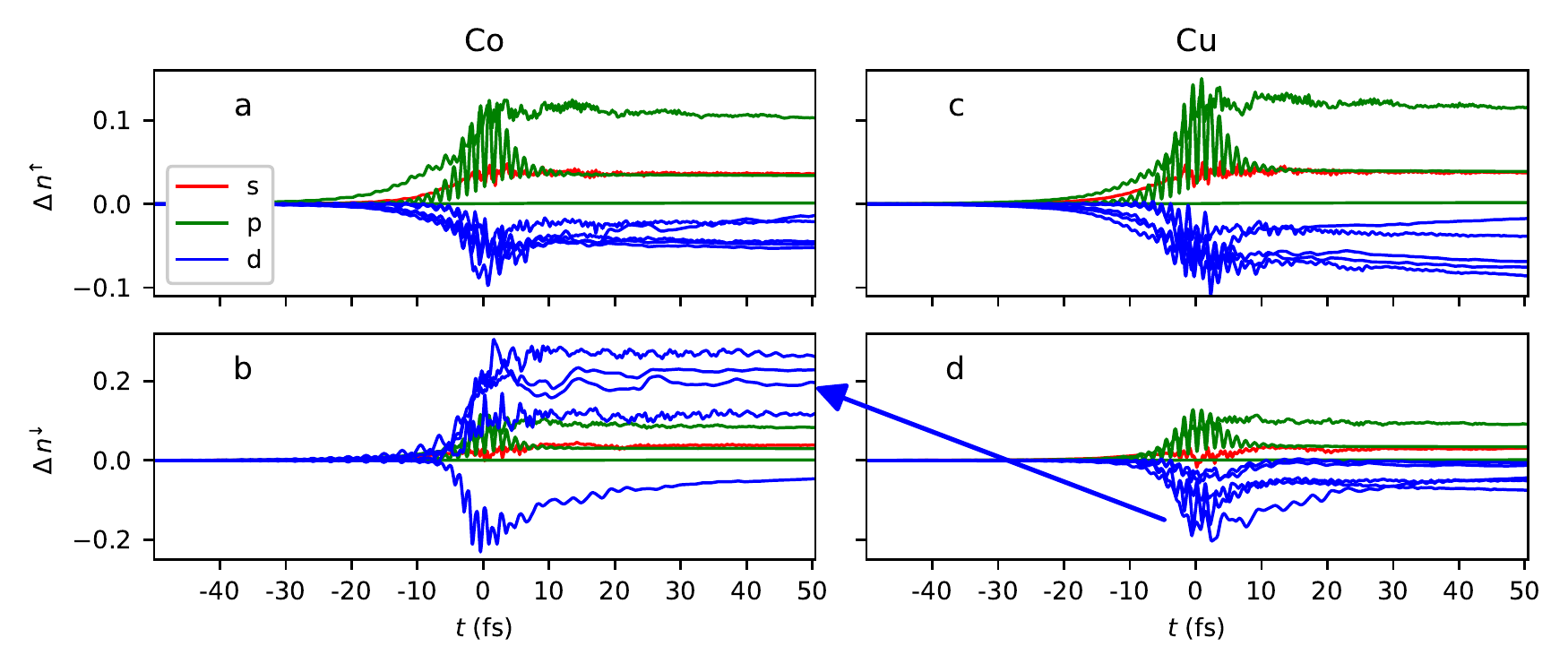}
    \caption{Time-dependent mean occupation numbers at the interface. Changes $\Delta n$ of occupation numbers of a Co site (left column, panels~a and ~b) and of a Cu site (right column, panels~c and~d) at the Co/Cu interface are resolved with respect to spin (top row: majority spin; bottom row: minority spin) and orbital angular momentum ($s$, $p$, and $d$: red, green, and blue, respectively). The blue arrow indicates the `reflow' associated with minority-spin $d$ orbitals across the interface (see text).}
    \label{fig:reflow}
\end{figure*}

The reflow is illustrated by means of occupation numbers of the interfacial sites. The spin-resolved occupation numbers of Co $sp$ and of Cu $sp$ orbitals are strongly modulated by the laser pulse (\Fref{fig:reflow}). A closer look reveals that simultaneously $p$ orbitals become populated, while $d$ orbitals become depopulated, which is readily explained by dipole transitions ($\delta l = -1$; both $p$ and $d$ occupation numbers exhibit similar slopes but with opposite sign). The $s$ orbitals become populated slightly after the $p$ orbitals because first $p$ orbitals have to be populated (before the laser pulse both $s$ and $p$ orbitals are weakly populated). 

The previous effect is contrasted by a depopulation of the $d$ orbitals, except for most of the Co $d^{\downarrow}$ orbitals; see panel~(b). The occupation numbers of one of the Co and Cu $d^{\downarrow}$ orbitals exhibit similar depletion shapes, which is interpreted by a dipole transition into respective $p$ orbitals. The remaining $d$ orbitals behave similarly but with Co populated and Cu depopulated: this observation suggest a spin-conserving `reflow' from Cu orbitals across the interface into these Co orbitals. Spin-resolved currents, discussed below, support this interpretation.

For the reflow effect it is crucial that the involved orbitals provide a sizable density of states: respective computations for aluminum (instead of copper) exhibit a strongly reduced effect, which is in line with the small density of states of aluminum in the relevant energy range.

The previous considerations corroborate the importance of interfaces for ultrafast spin transport. The local imbalance of occupation at an interface enhances the effect of the laser pulse; an interface may therefore be regarded as an additional  `source of disequilibrium' \footnote{We note in passing that simulations for homogeneous systems exhibit considerably less demagnetisation.}. The findings reported so far suggest also to distinguish two induced currents within the sample: a flow of hot, highly mobile $s$ and $p$ electrons contrasted by a reflow of  `tepid', weakly mobile $d$ electrons. These two contributions can be distinguished in space-time maps of charge and spin-resolved currents.

\begin{figure*}
    \centering
    \includegraphics[width = 0.48\columnwidth]{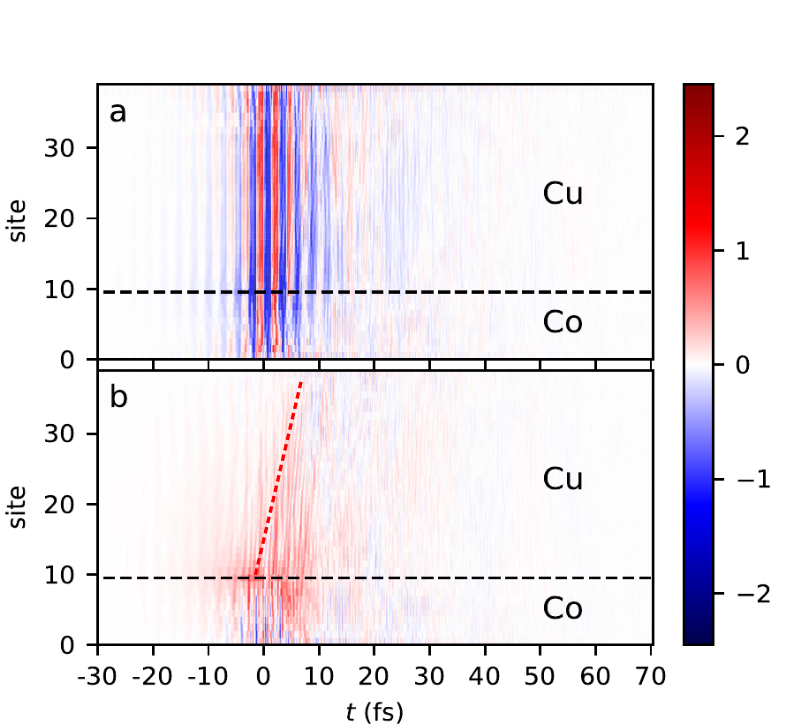}
    \nolinebreak
    \includegraphics[width = 0.48\columnwidth]{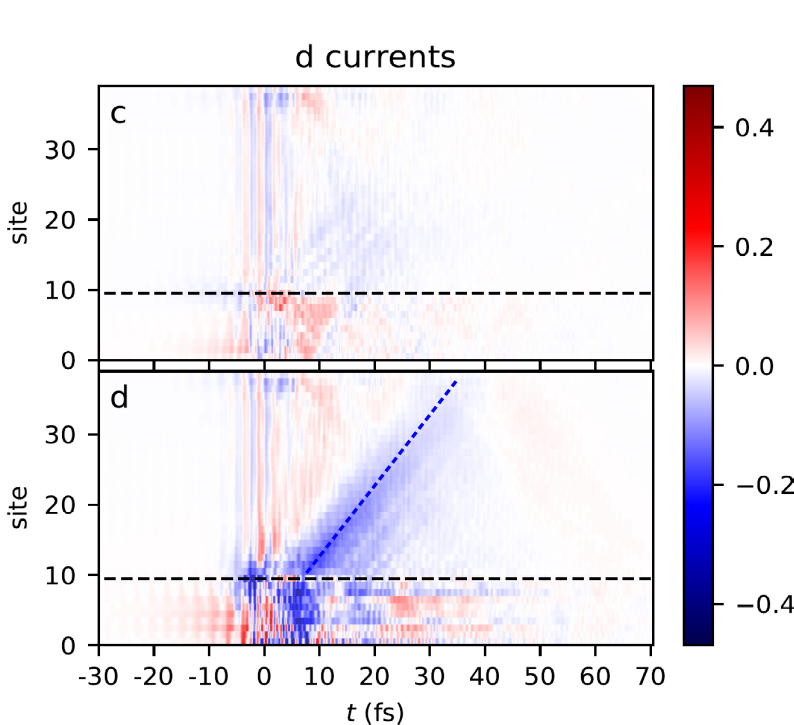}
    \caption{Spin- and orbital-resolved currents. Left: charge currents (a) and spin currents (b). Right: spin-resolved currents of $d$ orbitals only; majority spin (c) and minority spin (d). The color scales (one for each column; arbitrary units) indicate the transport direction: red (blue) toward the upper Cu edge (lower Co edge) of the sample. The colored dashed lines in panels~(b) and~(d) serve as guides to the eye for  estimating the velocities.}
    \label{fig:Currents}
\end{figure*}

The charge currents, defined in \Eref{eq:current}, are evenly distributed within the sample [\Fref{fig:Currents}(a)]. They are significantly modulated during the laser pulse; confer the alternating blue and red stripes at about $\unit[0]{fs}$. 

Regardless of the rather uniform distribution of the charge currents across the sample, spin is transferred from the Co region into the Cu region [panel~(b)]. The origin of the spin currents is clearly associated with the interface (dark red area around $\unit[0]{fs}$), from where it expands covering almost the entire sample. This expansion happens at a high velocity, as indicated by the dashed red line.

The redistribution of $d$ minority-spin electrons at the interface at $t \approx \unit[0]{fs}$ causes two phenomena. First, the currents in the Co regions are strongly modulated, which is explained by scattering of electrons at the sample's Co edge and at the interface [cf.\ the Co region in panel~(d) in the range from $\unit[5]{fs}$ to $\unit[40]{fs}$ with its intricate pattern]. Second, there is the reflow effect which shows up clearly as elongated blue area in the Cu region. It can be viewed as a depletion region of minority-spin $d$ electrons which, starting at the interface, `invades' the Cu region with low velocity (dashed blue line). It vanishes after about $\unit[40]{fs}$ due to dephasing. Moreover, this reflow -- figuratively speaking: a domino reaction -- contributes weakly to the total spin current shown in panel~(b); notice within this respect the different color scales for~(b) and~(d).

\section{Outlook} \label{sec:Outlook}
In this paper we have delineated that a density-operator approach based on a tight-binding description of the electronic structure is able to capture and reproduce essential phenomena of ultrafast spin dynamics and transport in inhomogeneous systems. To come closer to experiments, however, some areas need improvement. To name but a few: the coupling to the bath could include electron-phonon coupling strengths that are calculated for the actual sample (material) and the tight-binding parameters could be optimized for inhomogeneities. 

Concerning extensions of the approach, it is conceivable to incorporate electron-electron scattering and the electrostatic interaction, in particular screening, between regions with in- or decreased population (charge surplus or shortage). Noncollinear spin textures open up additional transport channels (that is spin-flipping hopping in the Hamiltonian $\hat{H}_{0}$) which could cause stronger demagnetisation \cite{Chen2019} and stronger spin currents; the textures may be either intrinsic (as in, e.\,g., Mn$_{3}$Sn~\cite{Tomiyoshi1982,Brown1990} and Mn$_{3}$Ir~\cite{Tomeno1999} or thermal (that is fluctuating magnetic moments, taken for example from atomistic spin dynamics simulations \cite{Erikssson2017}).

\ack
We acknowledge fruitful discussions with Sangeeta Sharma (Max Born Institute, Berlin). This work is supported by CRC/TRR 227 of Deutsche Forschungsgemeinschaft (DFG).

\appendix
\section{Protocol of a simulation} \label{sec:CompProc}
The method outlined in \Sref{sec:TheoreticalAspects} is implemented in our computer code \textsc{evolve} which is available from the authors.
A protocol of a typical simulation looks as follows.

\begin{enumerate}
    \item Setup a cluster in configuration space.
    
    \item Setup the Hamiltonian matrix $\mathsf{H}_{0}$ for the cluster in the site-orbital basis and compute its eigenvectors.
    
    \item  Determine the chemical potential $\mu$ and the initial occupation numbers $p_{i}^{(0)}$ (in the eigenstate basis) that enter the (diagonal) density matrix $\mathsf{P}^{(0)}$ using a Fermi-Dirac distribution for the prescribed temperature $T$ of the bath. \label{step:pre-therm}
    
    \item Starting with $\mathsf{P}^{(0)}$ and long before the laser pulse, evolve $\mathsf{P}(t)$ according to the Lindblad equation~\eref{eq:vonNeumann}. This differential equation is solved using an adaptive Runge-Kutta method of orders~2 and~3. An adaptive integration scheme proves advantageous since small time steps are indispensable during the laser pulse but redundant during relaxation. Data are usually written into a file every $\unit[0.1]{fs}$.
    \label{step:evolve}
\end{enumerate}

A typical simulation can be divided into three main stages: setup of cluster, Hamiltonian and initial occupation (step~\ref{step:pre-therm}), demagnetisation by the $\unit{fs}$ laser pulse, and subsequent relaxation toward thermal equilibrium on a $\unit{ps}$ timescale by coupling to the bath (step~\ref{step:evolve}). The bath is coupled to the electron system at all times.

\section{Evolution of mean occupation numbers} \label{sec:MeanOccuNum}
\begin{figure}[h]
    \centering
\includegraphics[width = 0.6\columnwidth]{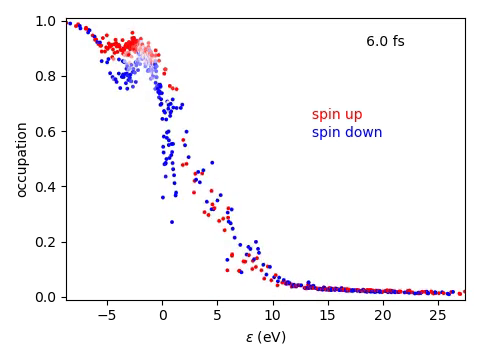}
    \caption{Energy- and spin-resolved mean occupation numbers of eigenstates of the Hamiltonian $\hat{H}_{0}$ in the time range $[-\unit[30]{fs} \ldots \unit[40]{fs}]$ (animation). Majority-spin (minority-spin) data are depicted as red (blue) dots. Parameters as for \Fref{fig:OccNumProfiles}.}
    \label{fig:Movie}
\end{figure}

The evolution of the mean occupation numbers is reproduced in \Fref{fig:Movie} as an animation. Snapshots of this animation taken for selected times are composed in \Fref{fig:OccNumProfiles} and discussed in \Sref{sec:OccNumDemag}.

\section{Spin-resolved currents}
\label{sec:spincurr}
Mahan expressed the current operator as \cite{mahan_many-particle_2000}
\begin{align*}
\vec{j} & = \mathrm{i} \sum_{kl} (\vec{r}_{k} - \vec{r}_{l}) \ket{k} h_{kl} \bra{l},
\end{align*}
in which the sums are over all sites within the sample and $h_{kl}$ is the respective tight-binding hopping element. This expression suggests to identify the current from site $l$ to site $k$ as
\begin{align*}
\vec{j}_{kl} 
& \equiv \frac{\mathrm{i}}{2}  (\vec{r}_{k} - \vec{r}_{l}) \left( \ket{k} h_{kl} \bra{l}
- \ket{l} h_{lk} \bra{k} \right).
\end{align*}
The flow direction is given by $\vec{r}_{k} - \vec{r}_{l}$, the current strength by
\begin{align*}
j_{kl} 
& \equiv \frac{\mathrm{i}}{2}  \left( \ket{k} h_{kl} \bra{l}
- \ket{l} h_{lk} \bra{k} \right),
\end{align*}
whose expectation value of $j_{kl}$ is
\begin{align*}
\braket{j_{kl}}  & = \frac{\mathrm{i}}{2} \left( \rho_{lk} \, h_{kl}  -  \rho_{kl} \, h_{lk} \right).
\end{align*}

In order to deduce spin-resolved currents, we extend the site indices to site, orbital, and spin: $k \mapsto (k, \alpha, \sigma)$ with $\sigma = \uparrow, \downarrow$ gives
\begin{align*}
\braket{j_{kl}} & =
\frac{\mathrm{i}}{2} 
\sum_{\alpha \sigma}   
\sum_{\beta \sigma'}   
\left[
\rho_{l \beta \sigma', k \alpha \sigma} \, h_{k \alpha \sigma, l \beta \sigma'}
-
\rho_{k \alpha \sigma, l \beta \sigma'} \, h_{l \beta \sigma', k \alpha \sigma} 
\right].
\end{align*}
Defining matrices $\mathsf{p}_{lk}^{\sigma \sigma'}$ and $\mathsf{t}_{lk}^{\sigma \sigma'}$ with elements
\begin{align*}
\left( \mathsf{p}_{lk}^{\sigma \sigma'} \right)_{\beta \alpha} & = \rho_{l \beta \sigma, k \alpha \sigma'},
\quad
\left( \mathsf{h}_{lk}^{\sigma \sigma'} \right)_{\beta \alpha} = h_{l \beta \sigma, k \alpha \sigma'},
\end{align*}
the expectation value reads
\begin{align*}
\braket{j_{kl}} & = 
\frac{\mathrm{i}}{2} \operatorname{tr}
\begin{pmatrix}
\mathsf{p}_{lk}^{\uparrow \uparrow} & \mathsf{p}_{lk}^{\uparrow \downarrow} \\
\mathsf{p}_{lk}^{\downarrow \uparrow} & \mathsf{p}_{lk}^{\downarrow \downarrow}
\end{pmatrix}
\begin{pmatrix}
\mathsf{h}_{kl}^{\uparrow \uparrow} & \mathsf{h}_{kl}^{\uparrow \downarrow} \\
\mathsf{h}_{kl}^{\downarrow \uparrow} & \mathsf{h}_{kl}^{\downarrow \downarrow}
\end{pmatrix}
- \left\{ l \leftrightarrow k \right\}.
\end{align*}
With
\begin{align*}
\mathsf{P}_{lk} & \equiv 
\begin{pmatrix}
\mathsf{p}_{lk}^{\uparrow \uparrow} & \mathsf{p}_{lk}^{\uparrow \downarrow} \\
\mathsf{p}_{lk}^{\downarrow \uparrow} & \mathsf{p}_{lk}^{\downarrow \downarrow}
\end{pmatrix},
\quad
\mathsf{H}_{kl} \equiv
\begin{pmatrix}
\mathsf{h}_{kl}^{\uparrow \uparrow} & \mathsf{h}_{kl}^{\uparrow \downarrow} \\
\mathsf{h}_{kl}^{\downarrow \uparrow} & \mathsf{h}_{kl}^{\downarrow \downarrow}
\end{pmatrix},
\end{align*}
the even more compact form
\begin{align*}
\braket{j_{kl}} & = 
\frac{\mathrm{i}}{2} \operatorname{tr}
\mathsf{P}_{lk} \mathsf{H}_{kl} - \left\{ l \leftrightarrow k \right\}
\end{align*}
is obtained.

The spin-polarised currents resolved with respect to the $\mu$-th component are given straightforwardly by
\begin{align*}
\braket{j_{kl}^{\mu}} & = 
\frac{\mathrm{i}}{4} \operatorname{tr}
\mathsf{P}_{lk}
\left[ \mathsf{\Sigma}^{\mu}, \mathsf{H}_{kl} \right]_{+}
- \left\{l \leftrightarrow k \right\}, \quad \mu = x, y, z,
\end{align*}
in which $\mathsf{\Sigma}^{\mu}$ is a block Pauli matrix and $[ \cdot,  \cdot]_{+}$ is the anti- or $+$-commutator.

\section*{References}
\bibliography{quellen}

\begin{thebibliography}{10}

\bibitem{krieger_laser-induced_2015}
K.~Krieger, J.~K. Dewhurst, P.~Elliott, S.~Sharma, and E.~K.~U. Gross.
\newblock Laser-{Induced} {Demagnetization} at {Ultrashort} {Time} {Scales}:
  {Predictions} of {TDDFT}.
\newblock {\em J. Chem. Theory Comput.}, 11(10):4870--4874, 2015.

\bibitem{Toews2014}
{W. T\"ows}.
\newblock {\em Many-body theory of laser-induced ultrafast demagnetization and
  angular momentum transfer in ferromagnetic transition metals}.
\newblock Phd thesis, University Kassel, Kassel, 2014.

\bibitem{Toews2015}
W.~T\"ows and G.~M. Pastor.
\newblock Many-body theory of ultrafast demagnetization and angular momentum
  transfer in ferromagnetic transition metals.
\newblock {\em Phys. Rev. Lett.}, 115:217204, Nov 2015.

\bibitem{koopmans_explaining_2010}
B.~Koopmans, G.~Malinowski, F.~Dalla~Longa, D.~Steiauf, M.~F\"ahnle, T.~Roth,
  M.~Cinchetti, and M.~Aeschlimann.
\newblock Explaining the paradoxical diversity of ultrafast laser-induced
  demagnetization.
\newblock {\em Nat. Mater.}, 9(3):259--265, 2010.

\bibitem{battiato_superdiffusive_2010}
M.~Battiato, K.~Carva, and P.~M. Oppeneer.
\newblock Superdiffusive {Spin} {Transport} as a {Mechanism} of {Ultrafast}
  {Demagnetization}.
\newblock {\em Phys. Rev. Lett.}, 105(2), 2010.

\bibitem{battiato_superdiffusive_2013}
M.~Battiato.
\newblock {\em Superdiffusive {Spin} {Transport} and {Ultrafast}
  {Magnetization} {Dynamics}: {Femtosecond} spin transport as the route to
  ultrafast spintronics}.
\newblock PhD thesis, University Uppsala, Uppsala, 2013.
\newblock OCLC: 940636548.

\bibitem{wieczorek_separation_2015}
J.~Wieczorek, A.~Eschenlohr, B.~Weidtmann, M.~Rösner, N.~Bergeard,
  A.~Tarasevitch, T.~O. Wehling, and U.~Bovensiepen.
\newblock Separation of ultrafast spin currents and spin-flip scattering in
  {Co}/{Cu}(001) driven by femtosecond laser excitation employing the complex
  magneto-optical {Kerr} effect.
\newblock {\em Phys. Rev. B}, 92(17), 2015.

\bibitem{Melnikov2011}
A.~Melnikov, I.~Razdolski, T.~O. Wehling, E.~Th. Papaioannou, V.~Roddatis,
  P.~Fumagalli, O.~Aktsipetrov, A.~I. Lichtenstein, and U.~Bovensiepen.
\newblock Ultrafast transport of laser-excited spin-polarized carriers in
  $\mathrm{Au}/\mathrm{Fe}/\mathrm{MgO}(001)$.
\newblock {\em Phys. Rev. Lett.}, 107:076601, 2011.

\bibitem{nenno_particle--cell_2018}
D.~M. Nenno, B.~Rethfeld, and H.~C. Schneider.
\newblock Particle-in-cell simulation of ultrafast hot-carrier transport in
  {Fe}/{Au} heterostructures.
\newblock {\em Phys. Rev. B}, 98(22):224416, 2018.

\bibitem{chen_competing_2019}
J.~Chen, U.~Bovensiepen, A.~Eschenlohr, T.~M\"uller, P.~Elliott, E.~K.~U.
  Gross, J.~K. Dewhurst, and S.~Sharma.
\newblock Competing {Spin} {Transfer} and {Dissipation} at {Co} / {Cu} ( 001 )
  {Interfaces} on {Femtosecond} {Timescales}.
\newblock {\em Phys. Rev. Lett.}, 122(6), 2019.

\bibitem{borchert_manipulation_2020}
M.~Borchert, C.~von Korff~Schmising, D.~Schick, D.~Engel, S.~Sharma, and
  S.~Eisebitt.
\newblock Manipulation of ultrafast demagnetization dynamics by optically
  induced intersite spin transfer in magnetic compounds with distinct density
  of states.
\newblock arXiv:2008.12612 [cond-mat], 2020.

\bibitem{dewhurst_laser-induced_2018}
J.~K. Dewhurst, P.~Elliott, S.~Shallcross, E.~K.~U. Gross, and S.~Sharma.
\newblock Laser-{Induced} {Intersite} {Spin} {Transfer}.
\newblock {\em Nano Lett.}, 18(3):1842--1848, 2018.

\bibitem{Slater1954}
J.~C. Slater and G.~F. Koster.
\newblock {Simplified LCAO Method for the Periodic Potential Problem}.
\newblock {\em Phys. Rev.}, 94:1498, Jun 1954.

\bibitem{Eich2017}
St. Eich, M.~Pl{\"o}tzing, M.~Rollinger, S.~Emmerich, R.~Adam, C.~Chen, H.~C.
  Kapteyn, M.~M. Murnane, L.~Plucinski, D.~Steil, B.~Stadtm{\"u}ller,
  M.~Cinchetti, M.~Aeschlimann, C.s~M. Schneider, and St. Mathias.
\newblock Band structure evolution during the ultrafast
  ferromagnetic-paramagnetic phase transition in cobalt.
\newblock {\em Sci. Adv.}, 3(3), 2017.

\bibitem{lindblad_generators_1976}
G.~Lindblad.
\newblock On the generators of quantum dynamical semigroups.
\newblock {\em Commun. Math. Phys.}, 48(2):119--130, 1976.

\bibitem{Pershin2008}
Yu.~V. Pershin, Y.~Dubi, and M.~Di~Ventra.
\newblock Effective single-particle {order-$N$} scheme for the dynamics of open
  noninteracting many-body systems.
\newblock {\em Phys. Rev. B}, 78:054302, 2008.

\bibitem{Coleman1960}
A.~J. Coleman.
\newblock {Structure of Fermion Density Matrices}.
\newblock {\em Rev. Mod. Phys.}, 32:175, 1960.

\bibitem{Chakraborty2014}
R.~Chakraborty and D.~A. Mazziotti.
\newblock {Generalized Pauli conditions on the spectra of one-electron reduced
  density matrices of atoms and molecules}.
\newblock {\em Phys. Rev. A}, 89:042505, 2014.

\bibitem{Ngyuen2015}
T.~S. Nguyen, R.~Nanguneri, and J.~Parkhill.
\newblock How electronic dynamics with {Pauli} exclusion produces {Fermi-Dirac}
  statistics.
\newblock {\em J. Chem. Phys}, 142:134113, 2015.

\bibitem{Head2015}
K.~Head-Marsden and D.~A. Mazziotti.
\newblock Satisfying fermionic statistics in the modeling of open
  time-dependent quantum systems with one-electron reduced density matrices.
\newblock {\em J. Chem. Phys}, 142:051102, 2015.

\bibitem{mahan_many-particle_2000}
G.~D. Mahan.
\newblock {\em Many-{Particle} {Physics}}.
\newblock Springer US, Boston, MA, 2000.

\bibitem{papaconstantopoulos_handbook_2015}
D.~A. Papaconstantopoulos.
\newblock {\em Handbook of the {Band} {Structure} of {Elemental} {Solids}}.
\newblock Springer US, Boston, MA, 2015.

\bibitem{Mathon1999}
J.~Mathon and A.~Umerski.
\newblock Theory of tunneling magnetoresistance in a junction with a
  nonmagnetic metallic interlayer.
\newblock {\em Phys. Rev. B}, 60:1117--1121, 1999.

\bibitem{alekhin_femtosecond_2017}
A.~Alekhin, I.~Razdolski, N.~Ilin, J.~P. Meyburg, D.~Diesing, V.~Roddatis,
  I.~Rungger, M.~Stamenova, S.~Sanvito, U.~Bovensiepen, and A.~Melnikov.
\newblock Femtosecond {Spin} {Current} {Pulses} {Generated} by the {Nonthermal}
  {Spin}-{Dependent} {Seebeck} {Effect} and {Interacting} with {Ferromagnets}
  in {Spin} {Valves}.
\newblock {\em Phys. Rev. Lett.}, 119(1), 2017.

\bibitem{Chen2019}
Z.~Chen and L.-W. Wang.
\newblock Role of initial magnetic disorder: {A} time-dependent ab initio study
  of ultrafast demagnetization mechanisms.
\newblock {\em Sci. Adv.}, 5:eaau8000, 2019.

\bibitem{Tomiyoshi1982}
S.~Tomiyoshi and Y.~Yamaguchi.
\newblock {Magnetic Structure and Weak Ferromagnetism of Mn$_{3}$Sn Studied by
  Polarized Neutron Diffraction}.
\newblock {\em J. Phys. Soc. Japan}, 51(8):2478--2486, 1982.

\bibitem{Brown1990}
P.~J. Brown, V.~Nunez, F.~Tasset, J.~B. Forsyth, and P.~Radhakrishna.
\newblock {Determination of the magnetic structure of Mn$_{3}$Sn using
  generalized neutron polarization analysis}.
\newblock {\em J. Phys. Condens. Matter}, 2(47):9409--9422, nov 1990.

\bibitem{Tomeno1999}
I.~Tomeno.
\newblock {Magnetic neutron scattering study of ordered Mn$_{3}$Ir}.
\newblock {\em J. Appl. Phys.}, 86:3853, 1999.

\bibitem{Erikssson2017}
O.~Eriksson, A.~Bergman, L.~Bergqvist, and J.~Hellsvik.
\newblock {\em {Atomistic Spin Dynamics: Foundations and Applications}}.
\newblock Oxford University Press, Oxford, 2017.

\end{thebibliography}
\bibliographystyle{unsrt}
\end{document}